# Development of New Hole-Type Avalanche Detectors and the First Results of their Applications

G. Charpak, P. Benaben, P. Breuil, A. Di Mauro, P. Martinengo, V. Peskov

*Abstract*-- We have developed a new detector of photons and charged particles- a hole-type structure with electrodes made of a double layered resistive material: a thin low resistive layer coated with a layer having a much higher resistivity. One of the unique features of this detector is its capability to operate at high gas gains (up to $10^4$) in air or in gas mixtures with air. They can also operate in a cascaded mode or be combined with other detectors, for example with GEM. This opens new avenues in their applications. Several prototypes of these devices based on new detectors and oriented on practical applications were developed and successfully tested: a detector of soft X-rays and alpha particles, a flame sensor, a detector of dangerous gases. All of these detectors could operate stably even in humid air and/or in dusty conditions. The main advantages of these detectors are their simplicity, low cost and high sensitivity. For example, due to the avalanche multiplication, the detectors of flames and dangerous gases have a sensitivity of 10-100 times higher than commercial devices. We therefore believe that new detectors will have a great future.

## I. INTRODUCTION

Hole-type gaseous detectors of photons and charged particles (capillary plate [1],GEM [2]) are very attractive in some applications due to their capability to operate at high gain in poorly quenched gases ( for example, in pure noble gases) and the capability to operate in cascade mode which allows to boost the overall gain. However, some of these devices, in particular GEM, are very fragile in handling and operation.

For the last several years we were focused on developing more robust version of hole- type electron amplifiers.

Our first attempt was to develop so called a Thick GEM (TGEM) [3, 4]. Further studies and developments of this promising device were performed later by Breskin group [5].

At the previous IEEE Nuclear Science Symposium we presented a new design: a TGEM, having electrodes coated with a resistive layer or fully made from a resistive material (for example Kapton 100XC10E5 [6]). We named this detector Resistive Electrode Thick GEM or RETGEM. This detector can operate at gain close to $10^5$ even in pure Ar and Ne and discharges at higher gain, due to the high resistivity of the electrodes, do not damage either the detector or the front-end electronics. Unfortunately, it turned out that it is not easy to obtain resistive Kapton from DuPont, due to some restrictions imposed to European users. For this reason we recently developed an alternative RETGEM manufacturing using the screen printing technology [7] which is used in microelectronics to produce patterns of different shape and resistivity and hence is widely available in many Labs and Companies. In all these previous RETGEM designs the HV to the detector electrode was applied via a Cu frame manufactured in the peripheral region of the detector (see Fig.1). This approach may cause problems in the case of large-area devices since the avalanche current should flow along the surface to the Cu frame thus producing undesirable voltage drop. A better solution for large-area detectors has been obtained developing a double resistive electrode RETGEM.

In this paper we will present first results of tests of this new device. In particular, the main focus of this report will be on a study of RETGEM operation in badly quenched gases, including ambient air, and new applications which such device offers.

## II. RETGEMs WITH DOUBLE LAYERED RESISTVE ELECTRODES

### A. A New RETGEM Design

As was already mentioned above, in the earlier RETGEM designs the HV to the detector electrodes was applied via the Cu frame manufactured in the peripheral area of the detector and the potential drop along the resistive surface created surface streamers [8].

In this work, to minimize the voltage drop along the surface we developed a double layered RETGEM prototype. The first step of its manufacturing was the same as described in [7]: a Cu frame was manufactured on the G-10 surface (the thickness of the G-10 plate was 0.5 or 1mm). The area inside the Cu frame (30x30 mm$^2$) was then coated by a vacuum evaporation technique with a 15 nm thick layer of Cr via a mask preventing the coating of 2D rows of circles, 0.8 mm in diameter - see Fig.2. The surface resistivity of this layer was ~50 KΩ/□. Afterwards, a resistive paste, Encre MINICO, was applied to the top and the bottom surfaces of the G-10 plate using screen printing technology. The paste is cured in air at 200° C for one hour. After the curing process is complete, the resistive layer was 50μm thick.

Holes with diameters of 0.3 or 0.5 mm were then drilled at even intervals (using a CNC machine) in the center of the circles free of Cr. The surface resistivity of the top layer was~ 0.5 M Ω/□. The schematic drawing of this double layered RETHGEM is shown in Fig. 3.

### B. Experimental Set up for the RETGEM Operation Study

Our experimental set up is schematically shown in Fig. 4. It consists of a gas chamber inside which a singe or double RETGEM can be installed, as well as the gas system and readout electronics. The distance between the drift mesh and the RETGEM can be varied from 1 to 4 cm. In some tests the voltage feeding of the RETGEM's electrodes was done using the resistive divider shown in Fig, 5. The gases used for initial tests were Ne, Ar, Ar+CO$_2$, however the main studies were done with pure air or mixtures of Ar with air. These tests were oriented towards new applications, described in the section III. The ionization inside the gas chamber was produced by alpha particles and 60 keV x-rays from $^{241}$Am and in some cases by 6 keV photons from $^{55}$Fe. If necessary, the active parts of the radioactive sources can be closed by shutters preventing the radiation to penetrate inside the chamber. The signals from the RETGEM's anode were detected by a CAEN charge-sensitive amplifier, and if necessary, further amplified by a research amplifier, treated by a LabView program and stored and on a PC.

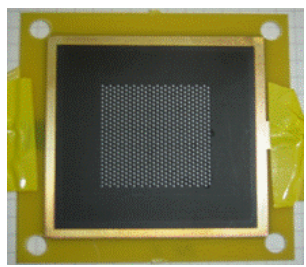

Fig. 1. A photo of the RETGEM (5x5cm$^2$) manufactured by the screen printing technology

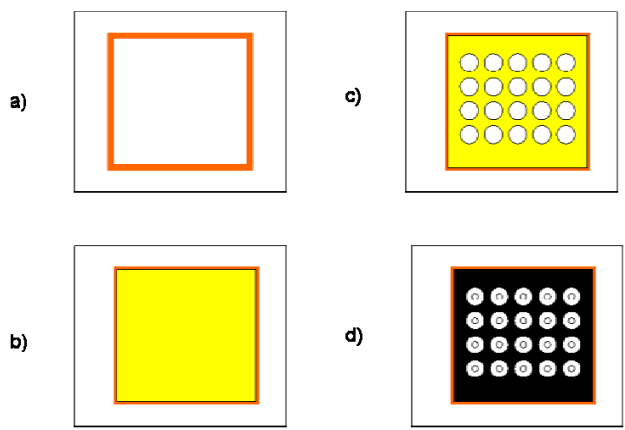

Fig. 2. A schematic drawing explaining four consequent steps in the double layered RETGEM manufacturing

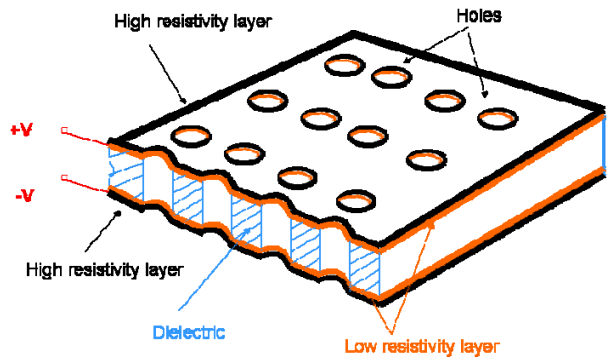

Fig. 3. A schematic drawing of the double layered RETGEM

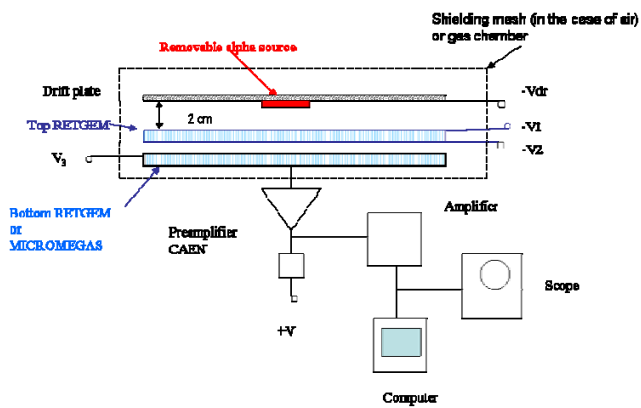

Fig. 4. A schematic drawing of the experimental set up for the study of single and double RETGEMs

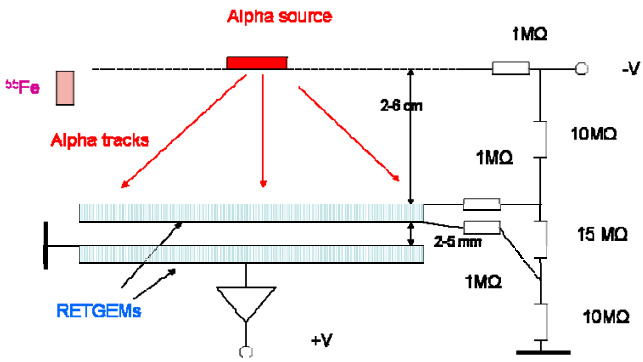

Fig. 5. A schematic drawing of the resistors divider chain used in some measurements with double RETGEMs.

## C. Results of RETGEM Tests

Fig. 6 shows a print out from the computer screen with a LabView image of results obtained with the RETGEM operating in Ar at 1 atm and detecting signals produced by alpha particles. The top screen shows the analogue signals from the RETGEM while the two screens on the bottom show the counting rate vs. time during these measurements (lower left screen) and their pulse-height spectrum (lower right screen). The activity of our Am source was measured with two independent dosimeters (Automess 6150 AD-k and $BaF_2$ scintillator coupled to a PM), the value of the alpha particles counting rate was $N_d$=100-110 c/s. As shown in Fig. 6, the alpha particles counting rate measured in Ar was 120c/s. It is slightly higher than those measured with the dosimeters, presumably due to the better collection of alpha track in the gas chamber compared to dosimeters where the tracks parallel to the detector surface were not detected. Hence, one can assume that ~100% efficiency was achieved in Ar.

Figs. 7 and 8 show signal amplitudes from the research amplifier (raw data obtained with alpha particles and with x-ray photons) vs. the voltage applied across the 1mm thick RETGEM -one with holes of 0.3 mm in diameter (Fig. 7) and the other one with holes of 0.5 mm in diameters operating in Ar or Ar+$CO_2$. From the data one can calculate the gas gain (our amplifiers were calibrated). For example, in the case of the Ar an Ar+$CO_2$ a 1V signal created by 6 keV photons corresponds to a gas gain of $10^3$.

During the study of the RETGEMs operation in badly quenched gases we have discovered that they can operate stably and at high gains not only in mixtures of Ar with air, but even in pure air. One should note that there were earlier attempts of various authors to investigate the operating in air of such "classical" detectors as wire -type or parallel plate type chambers, however it turned out that these detectors could not operate stably at gains larger than 10-100 [9]. There are several reasons for this, but the main one being the strong photon feedback preventing the achievement of higher gains.

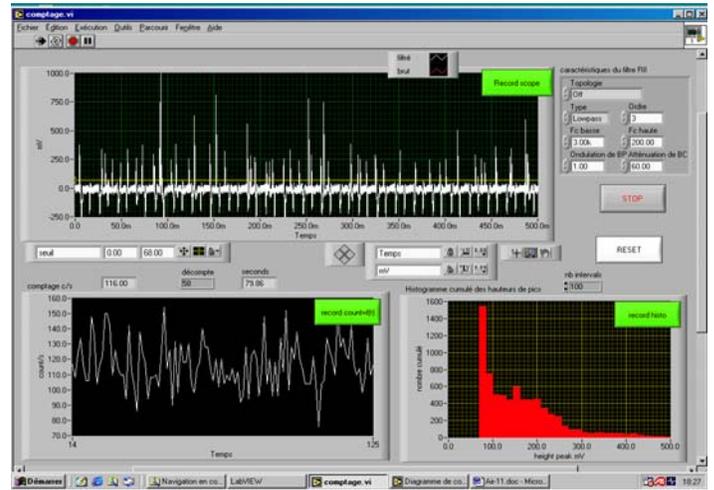

Fig. 6. A LabView screen showing the results obtained with RETGEM operating in Ar at a gain of 1.

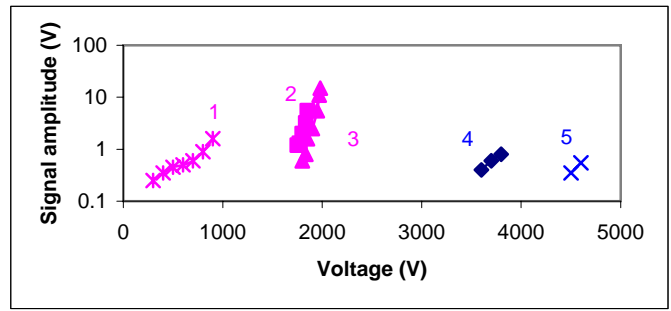

Fig. 7. Signal amplitude vs. voltage applied across RETGEM (1mm thick, holes 0.3 mm) operating in Ar (1, 2), Ar+$CO_2$ (3) and in air (4,5).1,4-signals were produced by alpha particles, 2,3-6keV, 5-60 keV

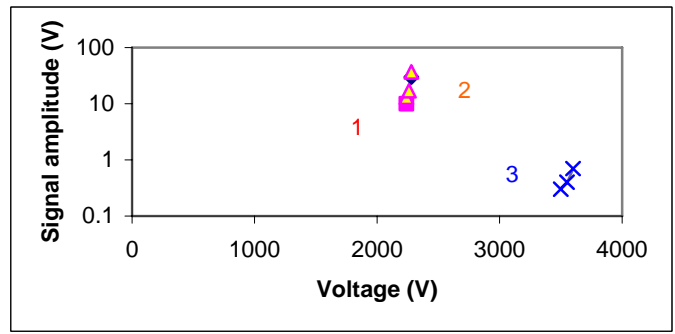

Fig. 8. Signal amplitude vs. voltage applied across RETGEM (1mm thick, holes 0.5mm) operating in Ar (1), Ar+$CO_2$ (2) and in air (3).1,2-6-keV, 3-alpha particles

One should add to this that in the case of discharges in the "classical" gaseous detectors, the energy released in the sparks depends on the gas, and it is very high in air which leads to the ordinary detectors and front end electronics to be damaged.

So why RETGEM can operate stably in air? In RETGEM, due to its hole -type geometry, the photon feedback is strongly suppressed (the cathode's electrode is geometrically shielded form the light emitted by the avalanches) and this allows to reach higher gains. Moreover, at high gain, like in resistive plate chambers (RPC) due to the charging up effect of the resistive surface around the holes, the electric field inside the holes diminishes for some short period of time, preventing the development of the successor avalanches according to the ion feedback mechanism. Thus at high gain both photon and ion feedbacks are strongly suppressed in the RETGEM. Finally, in case of the occasional discharges at very high gains, the RETGEM, like RPC, is spark-protected due to its dielectric electrodes.

In Fig. 9 are shown some results obtained with the RETGEM operating in pure air at high gas gain. The same counting rate was measured in air $N_{air}$ as was previously the case in Ar ($N_{Ar}$).

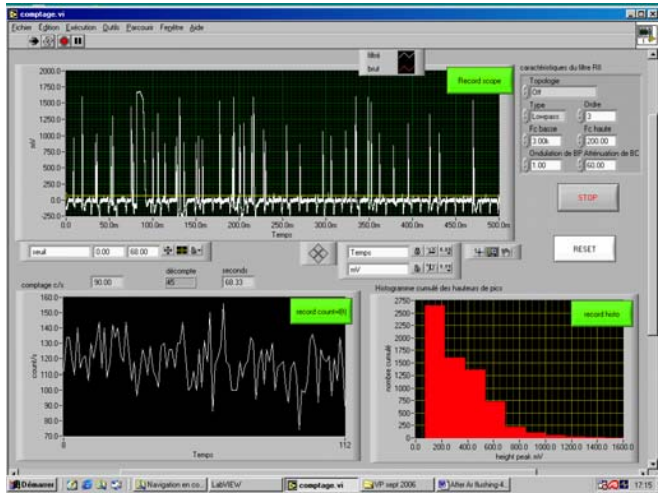

Fig. 9 A LabView screen showing the results obtained with the RETGEM operating in ambient air at humidity of 30%. The estimated gas gain ~$10^3$.

Thus the absolute efficiency for alpha detection in air is:
$$\varepsilon = N_{air}/N_{Ar} = 100\%. \quad (1)$$

In Figs. 7 and 8 in addition to the curves described above, are also presented the signal amplitudes vs. the voltage applied across the RETGEM operating in air. One can see that signals produced by alpha particles or 60 keV x-rays appeared in air at $V_{RETGEM} > 3500V$ and $V_{RETGEM} > 4500$ V respectively.

In the case of air from the measured amplitude of signals it is not easy to reliably calculate the gas gain because the number of primary electrons $n_0$ triggering the avalanches is not known. Indeed in air, most of the primary electrons produced by alpha particles or by X-ray photons are almost immediately attached to electronegative molecules and form electronegative ions. These ions drift to the detector's holes in which, in a strong electric filed, some experience the electron disattachment [10]. To our best knowledge it is not easy to calculate what fraction of electronegative ions lose their electrons. One can try to evaluate the $n_0$ from the efficiency measurements or from the energy resolution measurements, however this gives only a very rough estimation: $n_0$~10-100 - leading to gains being $10^4$-$10^3$, respectively.

The direct gain measurements were performed only with a photosensitive RETGEM (see section III-b) which allows us to independently confirm that at an applied voltage of ~3.5 kV across the RETGEM (0.5mm thick) the gain in air was ~ $10^4$. One should note that a single stage RETGEM operated stably only in dry air or when the humidity was below 30%. At a higher level of humidity it may exhibit some spurious noise pulses. On the other hands, double RETGEMs operate stably in air with humidity up to 80%. As an example, Fig. 10 shows results obtained with a double RETGEM in air at a humidity of 70%. An efficiency for alpha particle detection close to 100% was achieved and we thus consider this detector as more appropriate for practical applications.

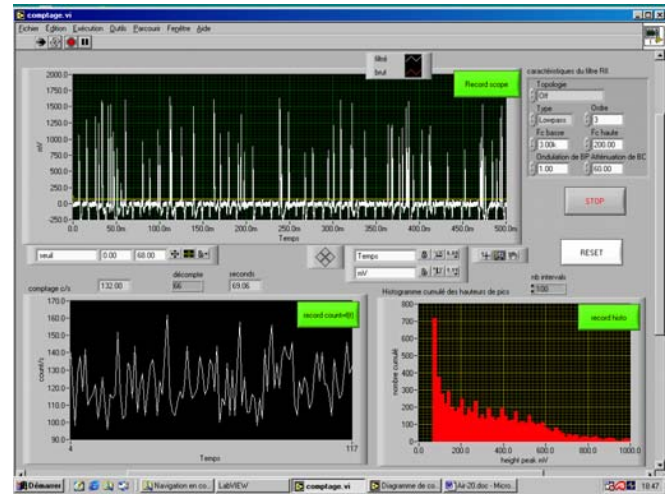

Fig. 10 A Labview screen showing the results obtained with double RETGEM operating in air at a humidity of 70%.

III. APPLICATIONS

RETGEMs operating in air in a cascaded mode may find wide range of applications. As examples, we will describe below prototypes of some practical devices which we have developed and tested recently: an alpha detector, a detector of flames and a detector of dangerous gases.

*A. Detection of Alpha Particles in Air*

The fact that double RETGEMs can detect alpha particles with ~ 100% efficiency makes them suitable as alpha particle background monitors which, due to their low cost can be used not only in houses, but in public areas.

A prototype of such a detector is shown in Fig. 11. It is an open-end metallic chamber inside which a double RETGEM is installed. Being placed ~5 mm apart from the surface containing an Am source it detects pulses produced by alpha particles (with counting rate $n_{air}$). To evaluate the detector efficiency the alpha particle counting rate emitted by the surface ($N_{dos}$) was measured with two standard dosimeters mentioned in the previous section: Automess 6150 AD-k and with a BaF$_2$ scintillator coupled to the PM.
The detection efficiency, defined as
$$\varepsilon = n_{air}/N_{dos} \quad (2)$$
was ~100%. Because the mean free path of alpha particles in air is ~4 cm, the detector could scan the surface containing alpha emitting elements on a distance larger than 0.5 cm, up to of 1-2 cm. The detection efficiency in this condition was 80-60% respectively. During these tests we have discovered that if the voltages $V_{dr}$, $V_1$, $V_2$, $V_3$ and $V_4$ are kept positively ( for example $V_{dr}$=2kv, $V_1$=4kV, $V_3$=5.5 , $V_4$=9kV ) it is possible to detect alpha particles with 60-70% efficiency even if the detector is placed 3-4 cm apart from the emitting surface. This is due to the fact that electronegative ions created near the grounded surface can drift towards the positively charged mesh and then at a proper voltage setting on the drift mesh and on the RETGEM's electrodes (as for example was mentioned above) a considerable fraction of them can be farther drifted towards the holes and trigger avalanches there. This feature makes our detector very different from any other existing alpha detectors (for example the dosimeters we used) in which alpha particles should hit the detector's sensitive area in order to be detected. In practice this limits the distance at which they can detect alpha particles to less than 2 cm. Of course, the energy resolution of our present prototype is much lower the best commercial devices, but due to its estimated low cost it can be massively used as stationary sensors in some areas which may require continuous monitoring of alpha particle contamination (for example Po): airports, railway stations and so on. In the present version, the RETGEM can be used as a trigger of the "first level" alarm in these areas, assuming that more refined analysis can be done a few minutes later with a more powerful and expensive portable alpha analyzer.

### B. Photosensitive RETGEM

We have already reported that RETGEMs coated with CsI photosensitive layers gain a high sensitivity to UV light [11]. Independent to these studies we also discovered that gaseous detectors combined with CsI photocathodes are very sensitive to the UV emission of flames and thus can be used as efficient flame sensors [12]. CsI photocathodes are rather robust, for example they can be exposed to ambient air for 10 min or so without destroying them. The main reason why the quantum efficiency (QE) degrades if CsI is exposed to ambient air for a longer period is the accumulation of water on the CsI surface, which changes its photoemission characteristics [13]. Thus one can assume that in dry air the CsI photocathode will be much more stable and if this is confirmed it opens the possibility to build cheap flame detectors operating in dry air.

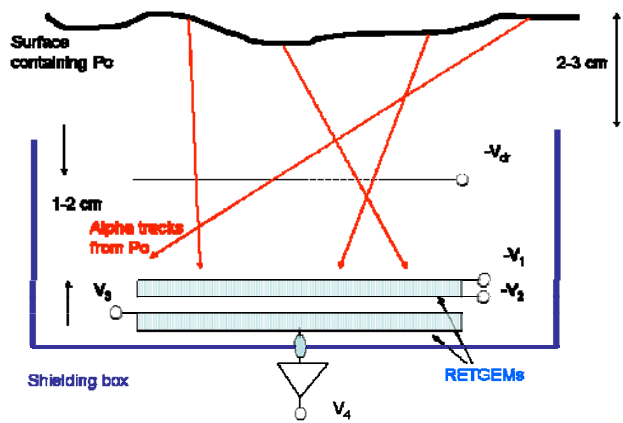

Fig. 11. A schematic drawing of a Po detector

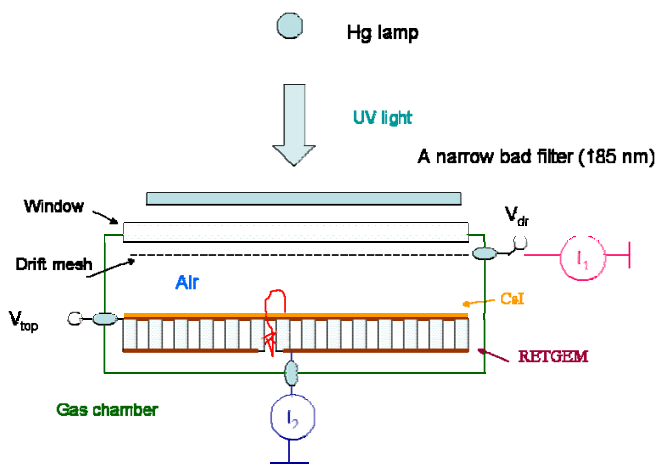

Fig. 12. A schematic drawing of the experimental set up for study of operation of RETGEMs combined with CsI photocathodes in air

To investigate the property of the CsI photocathode in air we used an experimental set up shown in Fig. 12. It contains a Hg UV lamp and a gas chamber with a CaF$_2$ window inside which a RETGEM was installed. The upper electrode of the RETGEM was coated be a vacuum evaporation technique with 0.35μm thick CsI layer. One cm above the RETGEM a drift mesh was installed. One could apply HV to the drift mesh or to the top electrode of the RETGEM and measure (with the help of a Kethley picoampermenter) the photocurrent produced by the Hg lamp either from the drift mesh ($I_1$) or from the RETGEM bottom electrode ($I_2$)-see Fig. 12.

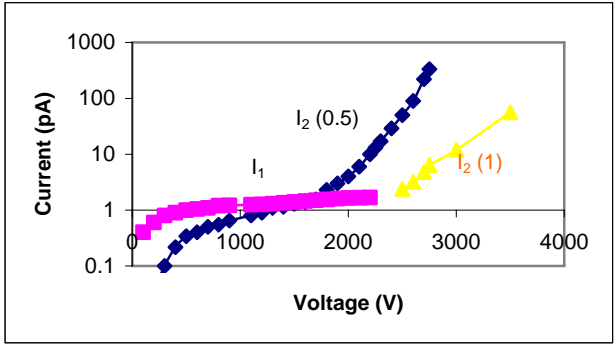

Fig.13. The photocurrent produced by an Hg lamp vs. the applied voltage measured from the drift mesh (the mesh was grounded via the picoampermeter and the negative voltage was applied to the RETGEM top electrode) and from the bottom electrode of the RETGEM (as function of the negative voltage applied to the top RETGEM electrode; in this measurement the voltage on the drift's electrode was kept constant and equal to -1 kV). The values in brackets indicate the RETGEM thickness

Some results are shown in Fig.13. The purple curve represents results of measurements of the photocurrent $I_1$ from the drift mesh produced by the Hg UV lamp as a function of the negative voltage applied to the top RETGEM's electrode $V_{top}$. One can see that this current first increases as a function of the voltage and then reaches a clear plateau, $I_{sat}$ at $V_{top} > 500V$ indicating that full collection of positive and electronegative ions is achieved. The blue curve represents the current $I_2$ measured from the bottom electrode of the RETGEM as a function of the negative voltage $V_{top}$ applied to the top electrode. The voltage at the drift's electrode in the measurements was $V_{drift}=1$ kV. One can see that at $V_{top}>1750V$ the current $I_g$ becomes higher than $I_{sat}$ indicating that RETGEM begin operating in a gas gain mode.

One can define the gas gain A as:
$$A = I_g / I_{sat} \quad (3)$$
and plot it as a function of the voltage applied across the RETGEM –see Fig. 14. If one extrapolates the gain curve to the voltages values obtained in the experiments described above with alpha particles and 60 keV x-ray photons in air (see section II-C) then we can conclude that the gain achieved in these measurements was $\sim 10^2$ -$10^3$ respectively.

From the absolute value of the $I_g$ (see Fig. 13) one can also estimate the CsI QE. To do this we performed additional calibration measurements with a reference detector - a single–wire counter filled with Ar+10%$CO_2$+TMAE, the QE of which is well known (see [14]). The depth of the active part of this detector was ~4 cm which ensures almost full absorption of the UV light from the Hg lamp. Fig. 15 shows the photocurrent vs. the applied voltage measured at wavelength $\lambda$=185 nm, with the reference detector and with the photosensitive RETGEM (see Fig. 12) operating

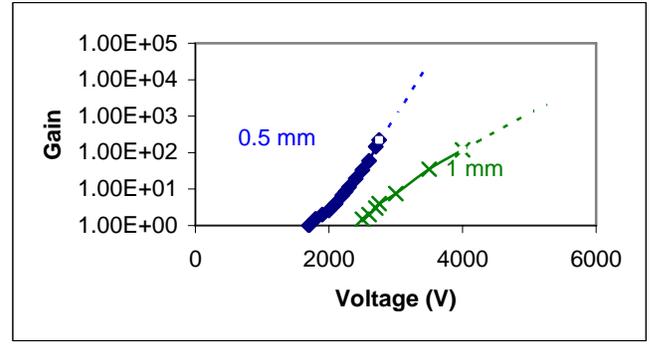

Fig. 14. Gain vs. voltage across RETGEM calculated from the data presented in Fig. 13.

in dry air. The RETGEM QE $Q_{RETGEM}$ can be calculated from the following equation:
$$Q_{RETGEM} = Q_{TMAE} I_g / I_{TMAE}, \quad (4),$$
where $Q_{TMAE}$ is TMAE QE at 185 nm ($Q_{TMAE}$=33%) and $I_{TMAE}$ is the saturated value of the photocurrent measured with the reference detector. As follows from Fig. 15, Ig=16 nA and $I_{TMAE}$=44 nA which give the value of $Q_{RETGEM}\sim 12$%.

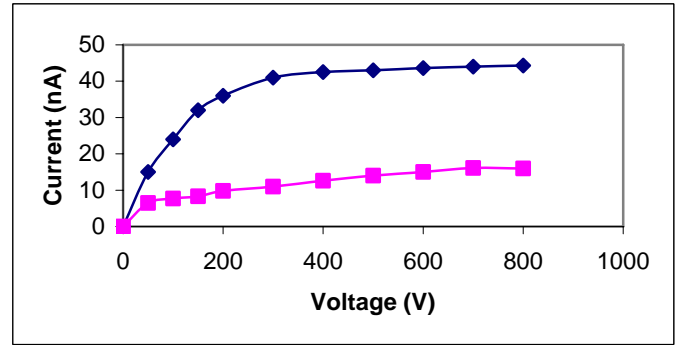

Fig. 15. Photocurrents vs. the voltage measured in the case of the reference detector (blue) and in the case of the photosensitive RETGEM with CsI photocathode operating in air (rose)

Fig. 16 shows $Q_{RETGEM}$ vs. time curves measured in ambient air and in dry air. It can be clearly seen that the QE in dry air is rather stable and this offers the possibility to build a rather cheap detector filled with dry air.

The prototype of such a detector built and tested is shown in Fig 17. It is a three-stage detector operating in dry air. It contains a double GEM combined with a RETGEM. The cathode of the upper GEM was coated with a CsI layer 0.4μm thick. The use of the GEMs in the first two stages of multiplication allows to minimize the effect of the photoelectron capturing by the electronegative molecules because the pitch of the holes in the GEM was only 140 μm and the distance between GEMs 0.5mm, so free primary electrons drift only short distances. The total multiplication in the double GEMs was ~100. The final stage is a robust spark-protected

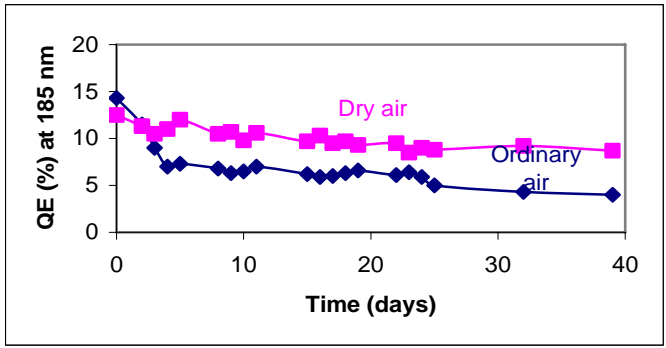

Fig.16. Stability measurements of CsI photocathodes in dry and ambient air.

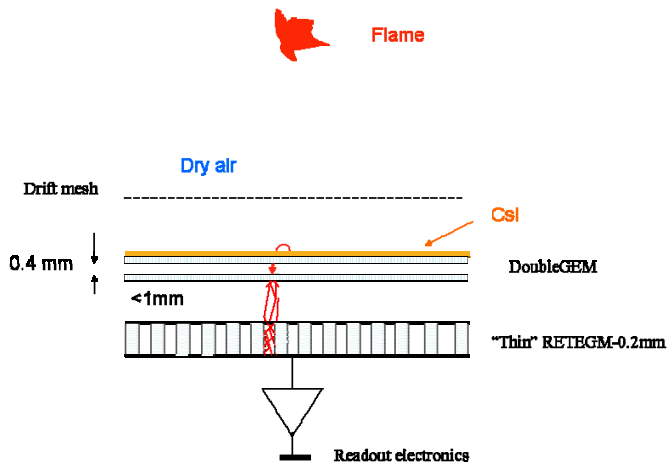

Fig.17. A schematic drawing of a UV flame detector operating in dry air

| Hamamatsu R2868 | | Our detector | | |
|---|---|---|---|---|
| Distance (m) | Mean number of counts per 10sec: $N_H$ | | Mean number of counts per 10sec: $N_O$ | Ratio $N_O/N_H$ |
| 3 | 76 | | 2034 | 26 |
| 10 | 6 | | 1820 | 30 |
| 15 | 2 | | 73 | ~35 |
| 30 | 0.1 | | 14 | N/A |

Table 1. Counting rate measured with R2868 and our detector both detecting a flame from candle placed

RETGEM- operated at estimated gain of $10^3$. This detector was able to detect single photoelectrons with a rather high efficiency. Table 1 shows counting rates measured with our detector and with a commercial UV flame detector Hamamatsu R2868, both detecting the same flame from a candle placed at different distances from the detector. From the presented data one can see that our detector is 25-30 times more sensitive than the Hamamatsu R2868 one. We are testing now a detector with larger sensitive area ($10 \times 10 cm^2$) and preliminary results indicate that its sensitivity is almost 85 times higher than Hamamatsu R2868.

One should note that strictly speaking there is no need to build a detector filled by dry air. This was just an extreme case to demonstrate the capability of our RETGEMs. In practice it is sufficient to manufacture a cheap," badly made" sealed detector (containing a water absorbing getter) filled with Ar or any other cheap gas and our result demonstrated that this in principle can be done.

### C. Detector of Dangerous Gases

Ionization chambers operating in air are widely used in various practical devices, for example as detectors of smoke [15] or dangerous gases (see for example [16]).

Certainly, in many applications a detector operating with gas gain may offer much higher sensitivity than ionization chambers. One of our projects was to develop a detector of dangerous gases based on the RETGEM.

A set up to study of a prototype of the RETGEM-based sensor of dangerous gases is shown in Fig. 18. It consist of a UV lamp, a testing gas chamber inside which a RETGEM detector was installed (a single- stage RETGEM and a drift mesh ~1cm above it), a pump system and a vessel filled with a liquid, the vapours of which the RETGEM is supposed to detect. The vessel was installed in a cryostat allowing to cool it in the range 300- 78K. This allows to introduce into the pumped testing chamber vapours of various liquids (ethylferrocene, benzene, toluene and so on) at a given temperature. For most of the liquids, the vapour pressure vs. the temperature is well known and thus we could introduce into the chamber a well known partial pressure of these liquids.

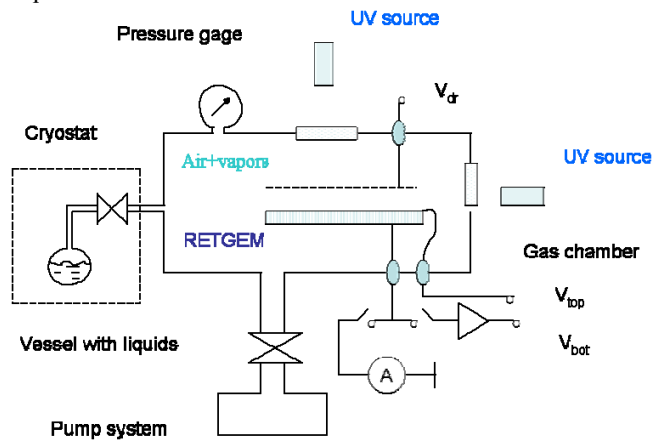

Fig. 18. A schematic drawing of the experimental set up for comparative studies of a conventional photoionization detector (ionization chamber) with a RETGEM- based detector of dangerous gases

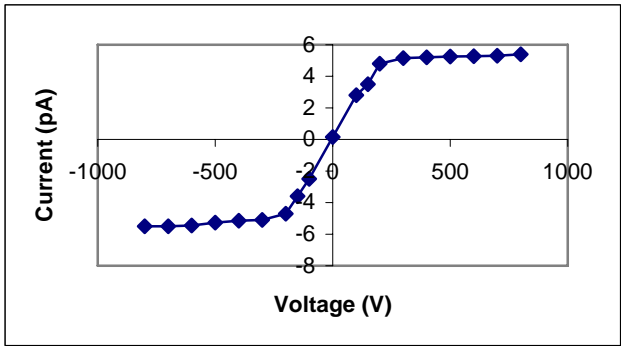

Fig.19. Photocurrent vs. the voltage between the drift mesh and the top RETGEM electrode in the case of the presence in air 1.3 ppm of ethylferrocene vapours

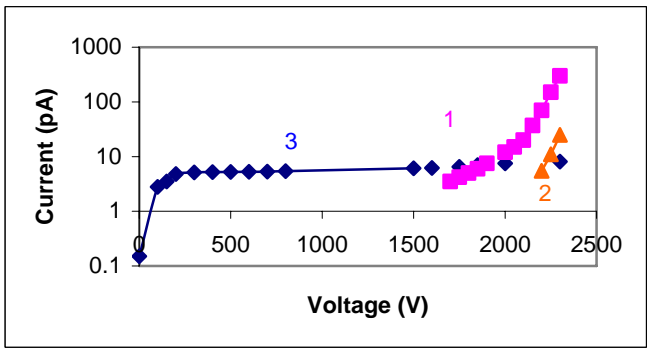

Fig. 20. Current measured from the RETGEM bottom electrode as function of the voltage across the RETGEM (The voltage drop between the drift mesh and the RETGEM top electrode was kept1kV) in the vase of 1.3 ppm (curve 1) and 0.1 ppm of ethylferrocene in air. Curve 3 shows results of measurements in ionization chamber mode (see Fig. 19)

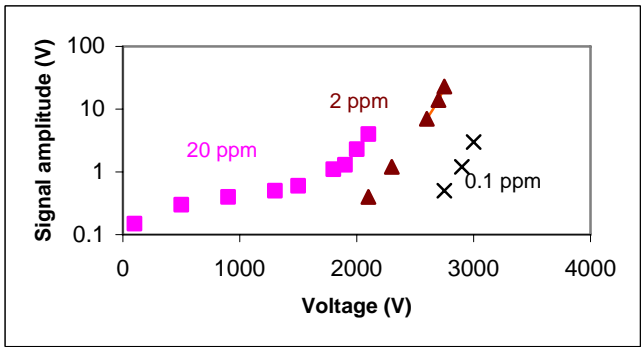

Fig. 21.Signal amplitudes from the RETGEM operating in pulse detection mode (Pulsed Ar lamp) as function of the voltage drop across the RETGEM for various concentrations of benzene in air

The testing chamber was then filled with ambient air. The UV light from the lamp, entering into the gap between the drift mesh and the RETGEM ionize the vapours and create a photocurrent. The RETGEM detector can operate in two modes: as a conventional ionization chamber and as a multiplication structure. In the first case the photocurrent was measured as a function of the voltage applied between the drift mesh and the top electrode of the RETGEM. As an example Fig. 19 shows the current-voltage (V-I) characteristics measured in air containing 1.3 ppm of ethylferrocene vapours. One can see that for both polarities of the voltage applied to the drift mesh the V-I curves are symmetrical and have clear current plateaus ($I_{sat}$) at high voltage. These results are very typical for photoionization detectors operating in air in the ionization chamber mode. The limit of their sensitivity to the UV source is determined by the minimum value of the photocurrent which can be reliably detected with a compact microelectronic circuit (usually ~pA). In the second case we measured the current from the bottom electrode of the RERGEM as a function of the voltage applied across its electrodes $\Delta V_{RETGEM}$ at a constant voltage drop between the drift mesh and the top electrode of the RETGEM. Some typical results obtained in this mode of operation are presented in the Fig. 20. One can see that the current increased with the $\Delta V_{RETGEM}$ and can finally exceed by one-two orders of magnitude $I_{sat}$. This allows to increase the sensitivity of the photodetector. Indeed, as shown in Fig. 20, with multiplication it is possible to detect 0.1ppm of ethylferrocene vapours, a level difficult to detect in the ionization chamber mode.

 We also performed some tests with a pulsed UV lamp. The advantage of the pulsed mode of operation is its much higher sensitivity than in a current mode because of the short current pulses which can be detected by a charge-sensitive amplifier. As an example Fig. 21 shows pulse amplitudes from the RETGEM (as a function of $\Delta V_{RETGEM}$) for various partial pressures of benzene down to 0.1 ppm. Note that this level of sensitivity is much higher than the limit typical limit of the commercially available photoionization detectors.
We thus we consider this development very promising

IV. CONCLUSIONS

 Ionization chambers working in ambient air in current detection mode are widely used in several applications such as smoke detection, dosimetry, therapeutic beam monitoring, dangerous gases detection and cetera [15-18]. In this work we demonstrated that hole-type structures can operate in air with gas gains. High gain and reliability can be achieved with RETGEMs which suppress not only photon feedback, but ion feedback as well and are spark-protected. In this paper we presented a new improved design of the RETGEM having double layer resistive electrodes which gives the possibility to building large-area, cost effective detectors. We also described prototypes and results of successful tests of several practical devices exploiting the gas multiplication in air: alpha detector, detector of flames and detector of dangerous gases. The main advantage of the alpha detector compared to commercial devices is its long-range capability: the alpha tracks can be detected 4-6 cm (and may be even more in improved designs) apart from the surfaces emitting alpha sources. The last two detectors, due to gas multiplication, have superior sensitivities to UV light than presently available commercial devices.

## V. Acknowledgments

We would like to thank A. Braem, R. Oliveira, J. Van Beelen, and M. Van Stenis for their help through out this work